\documentclass[12pt]{iopart}

\usepackage{bm}
\usepackage{graphicx}
\usepackage{color}
\usepackage{amssymb}
\usepackage[pdftex]{lscape}
\usepackage{slashbox}

\graphicspath{{img/}}

\begin{document}

\title[Low energy scattering in $e^-e^+\bar{p}$ and $e^+e^-\mbox{He}^{++}$ systems]
{High resolution calculation of low energy scattering in $e^-e^+\bar{p}$ and $e^+e^-\mbox{He}^{++}$ systems via  Faddeev-Merkuriev equations
%
%
}
\author{V A Gradusov, V A Roudnev, E A Yarevsky,
S L Yakovlev}
\address{Department of Computational Physics, St Petersburg State University, 7/9 Universitetskaya nab., St Petersburg, 199034, Russia}
\ead{s.yakovlev@spbu.ru}

\begin{abstract}

  The
  potential splitting approach incorporated into the framework of Faddeev-Merkuriev equations 
  in the differential form
  is used for calculations of multichannel scattering in  $e^-e^+\bar{p}$ and $e^+e^-\mbox{He}^{++}$ systems.
  Detailed calculations of all possible S-wave cross-sections are performed 
  in the low-energy region which { supports} 
  up to seven open channels including the rearrangement channels of ground and excited states of antihydrogen, positronium and helium ion formations.
  All known sharp resonances of the systems obtained and approved by a number of authors are clearly reproduced
  in the calculated cross sections. In cross sections for energies above the threshold corresponding to $n=2$ state of antihydrogen the prominent oscillations of Gailitis Damburg type have been found.
\end{abstract}
\submitto{\jpb}
\maketitle

\section{Introduction}

Study of electron and positron scattering off light atomic targets (like (anti)hydrogen atom and helium cation) is of  fundamental importance for atomic physics. These colliding systems represent genuine three-body Coulombic systems with variety of channels and with reach resonant structure of scattering cross sections.
Many calculations have been performed by now
for elastic and reactive scattering in these systems based on different computational platforms. Among others the variational methods \cite{bhatia74, humber97, humber01, gien99}, close coupling \cite{mitroy95, kadyr02} and
hyperspheric close coupling \cite{igara94, ward12} approaches and  methods of Faddeev-Merkuriev equations 
 \cite{kvits95, hu99, Papp-Yak2001, yak07, lazau17} 
 are in extensive use.
Other group of methods  exploit complex rotation of coordinates \cite{Res97} and complex rotation with potential splitting \cite{Volk09}, \cite{Yak10}, \cite{Volk11}, \cite{Volk15}, \cite{Yar15}, \cite{Yar17}. The latter is especially designed to treat the asymptotic Coulomb interaction and have been successfully applied for elastic electron (positron) hydrogen and electron (positron) helium cation collision.

Besides the features related to the long range character of the Coulomb interaction the collision in systems $e^+ - \mbox{H}$, $e^- - \bar{\mbox{H}}$ and $e^+ -\mbox{He}^+$ exhibits the fundamental rearrangement phenomenon of positronium
(electron-positron bound state) formation. For such a case the solution methods should be capable to efficiently represent the solution for all asymptotic fragmentations.
Highly accurate calculations of scattering processes and resonances in such systems of charged particles require {sophisticated}  methods from both theoretical and computational points of view.
The Faddeev equations \cite{fadd93} were designed especially to fulfill this requirement first for short-range interparticle interactions and then an extension on the long-range Coulomb case was made \cite{merkur80} . This extension, called the Faddeev-Merkuriev  (FM) equations, employs an idea of splitting the Coulomb potential into the interior and long range tail parts leading to the mathematically rigorous boundary value problem which solution is equivalent to the solution of the Schr\"odinger equation~\cite{fadd93}.
This approach perfectly suits  for the computationally difficult detailed low energy elastic and reactive scattering calculations in three-body Coulomb systems \cite{kvits92}, \cite{hu99}, \cite{Papp-Yak2001}.
Typically, splitting of the Coulomb potentials is done by an artificial functional parameter in the three-body configuration space, the Merkuriev cut-off function.
Recently, we showed in \cite{grad16}, that the splitting procedure can be made simpler and easier for implementing into FM equations.
 In this contribution we give practical advices for the choice of the cut-off function in the form of a smoothed Heaviside step function in the two-body configuration space, thus minimising the number of free parameters of the splitting procedure.
Following these advices allows us to produce more accurate results with smaller computational efforts.
Additionally, a special numerical technique for handling the matrices, known as ``tensor trick'', drastically reduces the computational complexity of the numerical solution of the FM equations~\cite{schelling89,roud00}.

Here, the formalism of FM equations has been used to calculate all possible S-wave cross sections in $e^-e^+\bar{p}$ and $e^+e^-\mbox{He}^{++}$ systems in the low-energy region.
Even though there are many calculations available in the literature, there is some lack of high-precision and detailed results.
A special emphasis is made on antihydrogen formation by antiproton impact of positronium which is currently used in experiments on antimatter at CERN~\cite{lazau18} (and references therein).

The paper is organised as follows. In section 2 we give necessary portion of the three-body FM equations formalism with splitting of the long-range Coulomb potential. Section 3 describes the solution technique of the resulting FM equations in the total angular momentum representation in 3D configuration space. Section 4 contains results of calculations of low-energy reactive scattering in $e^-e^+{\bar{p}}$ and $e^+e^-\mbox{He}^{++}$ systems. Last section 5 concludes the paper.

Throughout the paper we use atomic units.

\section{Theory}

We consider the system of three spinless nonrelativistic charged particles of masses $m_{\alpha}$ and charges $Z_\alpha$, $\alpha=1,2,3$. In what follows the set of indices $\{\alpha$, $\beta$, $\gamma\}$ runs over the set $\{1,2,3\}$ enumerating particles and it is also used for identifying the complementary pair of particles,
since
in the partition $\{\alpha(\beta \gamma)\}$ the pair of particles $\beta\gamma$ is uniquely determined by the particle $\alpha$.
The standard Jacobi coordinates are defined for a partition $\alpha(\beta\gamma)$ as relative position vectors between the particles of the pair $\beta\gamma$ and between their center of mass and the particle $\alpha$.
In applications it is convenient to use reduced Jacobi coordinates $\bm{x}_\alpha$, $\bm{y}_\alpha$ which are Jacobi vectors scaled by factors $\sqrt{2\mu_{\alpha}}$ and $\sqrt{2\mu_{\alpha(\beta\gamma)}}$, respectively, where the reduced masses are given by
\begin{equation}
\mu_{\alpha} = \frac{m_\beta m_\gamma}{m_\beta+m_\gamma},\quad \mu_{\alpha(\beta\gamma)}=\frac{m_\alpha (m_\beta+m_\gamma)}{m_\alpha+m_\beta+m_\gamma}.
\end{equation}
For different $\alpha's$ the reduced Jacobi vectors are related by an orthogonal transform
\begin{equation}
\bm{x}_\beta=c_{\beta\alpha}\bm{x}_\alpha + s_{\beta\alpha}\bm{y}_\alpha \ \
\bm{y}_\beta=-s_{\beta\alpha}\bm{x}_\alpha + c_{\beta\alpha}\bm{y}_\alpha,
\label{JacobiTrans}
\end{equation}
where
$$
c_{\beta\alpha}=-\left[\frac{m_\beta m_\alpha}{(M-m_\beta)(M-m_\alpha)}\right]^{1/2} \ \
s_{\beta\alpha}=(-1)^{\beta-\alpha}\mbox{sgn}(\alpha-\beta)(1-c^2_{\beta\alpha})^{1/2}
$$
and $M=\sum_\alpha m_\alpha$.
In what follows where it is due, it is assumed that $\beta$ Jacobi vectors are represented through $\alpha$ vectors via (\ref{JacobiTrans}).

In these reduced Jacobi coordinates FM equations for three charged particles \cite{fadd93} read
\begin{eqnarray}
\label{MFeq}
\{ T_\alpha  +  V_\alpha(x_\alpha) +
\sum_{\beta\ne\alpha}V_\beta^{(l)}(x_\beta,y_\beta) &-& E \} \psi_\alpha(\bm{x_\alpha}, \bm{y_\alpha}) = \nonumber \\
  &-&  V_\alpha^{(s)}(x_\alpha,y_\alpha) \sum_{\beta\ne\alpha}\psi_\beta(\bm{x_\beta}, \bm{y_\beta}).
\end{eqnarray}
Here $T_\alpha\equiv-\Delta_{\bm{x_\alpha}} - \Delta_{\bm{y_\alpha}}$ are the kinetic energy operators. Throughout the paper the magnitude of a vector $\bm{x}$ is denoted by $x$, i.e. $x=|\bm{x}|$, for a unit vector $\bm{x}/x$ the notation $\hat{\bm{x}}$ is used.
The potentials $V_\alpha$ represent the pairwise Coulomb interaction $V_{\alpha}(x_\alpha)=\sqrt{2\mu_{\alpha}}Z_\beta Z_\gamma/x_\alpha$ ($\beta,\gamma\ne\alpha$), although a short-range (decreasing as $1/x^2_{\alpha}$ or faster as $x_\alpha \to \infty$) potential can also be included in the formalism. For definitiveness and in view of further application of the formalism to the concrete systems of this paper we assume that one of the Coulomb potentials is repulsive whereas the other two are attractive.
We chose assignments such that the inequalities $Z_1Z_2>0$, $Z_1Z_3<0$ and $Z_2Z_3<0$ hold,
thus $V_3$ is always  repulsive.
The potentials $V_\alpha$ are split into interior (short-range) $V^{(s)}_\alpha$ and tail (long-range) parts $V^{(l)}_\alpha$
\begin{equation}
\label{PotSplit}
V_\alpha(x_\alpha) = V^{(s)}_\alpha(x_\alpha,y_\alpha) + V^{(l)}_\alpha(x_\alpha, y_\alpha).
\end{equation}
The equations~(\ref{MFeq}) can be summed up leading to the Schr\"odinger equation for the wave-function
$\Psi=\sum_{\alpha}\psi_\alpha$, wherein functions $\psi_{\alpha}$ are called the components of the wave function.

Splitting (\ref{PotSplit}) of the potentials in general case is done in the three-body configuration space by the  Merkuriev cut-off function $\chi_\alpha$ \cite{fadd93}
\begin{equation}
V_\alpha^{(s)}(x_\alpha,y_\alpha) = \chi_\alpha(x_\alpha, y_\alpha) V_\alpha(x_\alpha).
\label{split3b}
\end{equation}
 This splitting  confines  the short-range part of the potential to the regions in the three-body configuration space corresponding to the three-body collision point (particles are close to each other) and the binary configuration ($x_\alpha \ll y_\alpha$, when $y_\alpha \to \infty$). The form of the cut-off function can be rather arbitrary within some general requirements~\cite{merkur80}. Traditionally, the following form of the cut-off function proposed in~\cite{kvits92} is in use
\begin{equation}
\label{Mcutoff0}
\chi_\alpha(x_\alpha, y_\alpha) =
2/\left\{1+\exp[ (x_\alpha/x_{0\alpha})^{\nu_\alpha}/(1+y_\alpha/y_{0\alpha}) ]\right\}.
\end{equation}
The parameters $x_{0\alpha}$ and $y_{0\alpha}$ can in principle be chosen arbitrarily, but their choice changes the properties of components $\psi_\alpha$ important from both the theoretical and computational points of view \cite{YakPapp2010}.
The splitting procedure (\ref{PotSplit},\ref{Mcutoff0}) is done in the three-body configuration space and is suitable for energies as below the three-body breakup threshold as well as above this threshold.
In the paper \cite{grad16} we have shown, that for energies below the disintegration threshold splitting
can be vastly simplified by confining a cut-off function $\chi_\alpha(x_\alpha)$ on the two-body configuration space.
Formally it is obtained by chousing $y_{0\alpha}=\infty$,  which leads to
\begin{equation}
\label{Mcutoff}
\chi_\alpha(x_\alpha) = 2/\left\{1+\exp[ (x_\alpha/x_{0\alpha})^{\nu_\alpha} ]\right\}.
\end{equation}
With this smoothed Heaviside step function with $\nu_\alpha=2.01$ which is used in this paper for actual calculations, the splitted potentials $V^{(s,l)}_\alpha$
become the two-body quantities $V^{(s,l)}_\alpha=V^{(s,l)}_\alpha({x}_\alpha)$.

The splitting procedure makes the properties of the FM equations for Coulomb potentials for treating the scattering problem as appropriate as standard Faddeev equations in the case of short-range potentials \cite{Papp-Yak2001}. The key one of these properties  of FM equations~(\ref{MFeq}) 
is that the right-hand side of each equation is localized in the region of configuration space corresponding to three-body collision point \cite{YakPapp2010}. It results in the asymptotic uncoupling of the
set of FM equations and, accordingly, the asymptote of each component $\psi_\alpha$ for energies below the breakup threshold contains only terms corresponding to binary configurations of pairing $\alpha$  \cite{YakPapp2010, Papp-Yak2001}.
For the total energy $E$ of the system below the three-body ionization threshold it reads
\begin{eqnarray}
\label{asympt3D}
\psi_\alpha(\bm{x_\alpha},\bm{y_\alpha}) & = & \Phi^{\mathrm{in}}_{A_0}(\bm{x}_\alpha,\bm{y}_\alpha;\bm{p}_{n_0}) \delta_{\alpha,\alpha_0} \nonumber \\
& + &\sum\limits_{A}\frac{\phi_{n\ell}(x_\alpha)}{x_\alpha}Y_{\ell m}(\hat{\bm{x}}_\alpha)\sqrt{\frac{p_{n_0}}{p_n}}
\mathcal{A}_{A,A_0}(\hat{\bm{y}}_\alpha)
\frac{e^{i(p_n y_\alpha-\eta_n\ln(2p_n y_\alpha))}}{y_\alpha},
\end{eqnarray}
where the multi-index $A=\{n\ell m\}$ specifies various two-body Coulombic bound states of particles of pairing $\alpha$ (that is, binary scattering channels $\{\alpha;A\}$) with wave function $\phi_{n\ell}(x_\alpha)Y_{\ell m}(\hat{\bm{x}}_\alpha)/x_\alpha$ and energy $\varepsilon_n$. Here and in what follows by $Y_{\ell m}(\hat{\bm x})$ the standard spherical harmonic is denoted.  The momentum $p_n$ of the outgoing particle 
is determined by the energy conservation condition $E=p_n^2+\varepsilon_n$ and the Sommerfeld parameter is defined as  $\eta_n\equiv Z_\alpha(\sum_{\beta\ne\alpha}Z_\beta)\sqrt{2m_{\alpha(\beta\gamma)}}/(2p_n)$. The binary scattering amplitude $\mathcal{A}_{A,A_0}(\hat{\bm{y_\alpha}})$ corresponds to transition from the initial binary channel $\{\alpha_0;A_0\}$ to the binary channel $\{\alpha;A\}$. The initial channel is specified by the incoming wave, generated
by the asymptotic Coulomb interaction between the projectile and the target
 $V_{\alpha}^{\mathrm{eff}}(y_\alpha)=2p_n\eta_n/y_\alpha$ ($\beta,\gamma\ne\alpha$)
\begin{eqnarray}
\Phi^{\mathrm{in}}_{A_0}(\bm{x}_\alpha,\bm{y}_\alpha;\bm{p}_{n_0}) & = & \frac{\phi_{n_0\ell_0}(x_\alpha)}{x_\alpha}Y_{\ell_0m_0}(\hat{\bm{x}}_\alpha)
e^{i(\bm{p_{n_0}},\bm{y_\alpha})} e^{-\pi\eta_{n_0}/2}\Gamma(1+i\eta_{n_0}) \nonumber \\
& \times & {}_1F_1(-i\eta_{n_0}, 1, i(p_{n_0}y_\alpha-(\bm{p_{n_0}},\bm{y_\alpha}))),
\end{eqnarray}
where ${}_1F_1$ is the confluent hypergeometric function \cite{abram86}.

The total angular momentum is an integral of motion for the three-particle system. This fact allows one to reduce
the set of FM equations by projecting (\ref{MFeq}) onto a subspace of a given total angular momentum
\cite{kvits95}.
In this article we consider the case of zero total angular momentum of the system.
The kinetic energy operator in the left-hand side of equations~(\ref{MFeq}) on the subspace of the zero
total orbital momentum have the form
\begin{eqnarray}
T_\alpha = & - & \frac{\partial^2}{\partial y_\alpha^2} - \frac{2}{y_\alpha}\frac{\partial}{\partial y_\alpha}
- \frac{\partial^2}{\partial x_\alpha^2} - \frac{2}{x_\alpha}\frac{\partial}{\partial x_\alpha} \nonumber \\
& - & \left(\frac{1}{y_\alpha^2}+\frac{1}{x_\alpha^2}\right)
\frac{\partial}{\partial z_\alpha}(1-z_\alpha^2)\frac{\partial}{\partial z_\alpha},
\end{eqnarray}
where $z_\alpha\equiv\cos(\hat{\bm{x}}_\alpha\cdot\hat{\bm{y}}_\alpha)$. The corresponding projection of the
component $\psi_\alpha$ depends only on the coordinates $X_\alpha=\{x_\alpha,y_\alpha,z_\alpha\}$ in the plane containing all three particles.
By choosing appropriately the coordinate system the projection of the asymptote~(\ref{asympt3D}) on the state with zero total angular momentum can be written as (the constant factor is omitted)
\begin{eqnarray}
\label{zAs}
\psi_\alpha(X_\alpha) & \sim & -\frac{\phi_{n_0\ell_0}(x_\alpha)}{x_\alpha y_\alpha}
Y_{\ell_0 0}(\theta_\alpha,0)
 e^{-i\vartheta_{\ell_0}(y_\alpha,p_{n_0})} \delta_{\alpha,\alpha_0}\nonumber\\
& + & \sum\limits_{n\ell}
\frac{\phi_{n\ell}(x_\alpha)}{x_\alpha y_\alpha}
Y_{\ell 0}(\theta_\alpha,0)
\sqrt{\frac{p_{n_0}}{p_n}}
S_{n\ell,n_0\ell_0}
e^{+i\vartheta_{\ell}(y_\alpha,p_n)},
\end{eqnarray}
where $\vartheta_{\ell}(y_\alpha,p_n)\equiv p_{n}y_\alpha-\eta_{n}\ln(2p_{n}y_\alpha)-\ell\pi/2+\sigma_{n}$, $\sigma_{n}=\arg\Gamma(1+i\eta_{n})$ and $S_{n\ell,n_0\ell_0}=
\delta_{n\ell,n_0\ell_0}
+2ip_{n_0}e^{-i(\sigma_n+\sigma_{n_0})}e^{i\pi(\ell-\ell_0)/2}A_{n\ell,n_0\ell_0}$ are the $S$-matrix elements.
The total amplitude $\mathcal{A}_{AA_0}$ is connected to partial amplitudes $A_{n\ell,n_0\ell_0}$ according to
\begin{equation}
\sum_{m=-\ell}^{\ell}\int\mathrm{d}\hat{\bm{y_\alpha}}|\mathcal{A}_{AA_0}(\hat{\bm{y}}_\alpha)|^2 = 4\pi|A_{n\ell,n_0\ell_0}|^2+\ldots,
\end{equation}
where the omitted terms in the right-hand side are contributions from higher total angular momenta.

\section{Numerical solution}

To reduce the computational cost of solving the system of FM equations~(\ref{MFeq}), few modifications are  made. At first, since the potential $V_3$ is repulsive and thus the two-body Hamiltonian with this potential does not support bound states, this potential is included in the left-hand side of equations~(\ref{MFeq}) thus reducing the number of equations from 3 to 2:
\begin{equation}
\label{2FM}
\{T_\alpha + V_\alpha(x_\alpha) +
V_\beta^{(l)}(x_\beta) + V_3(x_3) - E\} \psi_\alpha(X_\alpha) =
 -V_\alpha^{(s)}(x_\alpha) \psi_\beta(X_\beta),
\end{equation}
where $\beta\ne\alpha=1,2$.
Formally, it is done by setting  $\chi_3=0$. 
Secondly, the asymptotic particle-atom Coulomb potential
$V_{\alpha}^{\mathrm{eff}}(y_\alpha)$ 
is introduced explicitly  in~(\ref{2FM}) for treating the asymptotic Coulomb singularity
\begin{eqnarray}
\{T_\alpha  +  V_\alpha(x_\alpha) & + &
V_\alpha^{\mathrm{eff}}(y_\alpha) - E\} \psi_\alpha(X_\alpha) =
  -  V_\alpha^{(s)}(x_\alpha) \psi_\beta(X_\beta) \nonumber \\
 & - & \left[ V_\beta^{(l)}(x_\beta) + V_3(x_3) - V_\alpha^{\mathrm{eff}}(y_\alpha) \right]\psi_\alpha(X_\alpha).
\end{eqnarray}
After that the Coulomb singularity can be effectively inverted as described below.
Thirdly, it is more convenient to work with the modified components
\begin{equation}
\widetilde{\psi_\alpha}(X_\alpha) = x_\alpha y_\alpha \psi_\alpha(X_\alpha)
\end{equation}
that satisfy the system of equations
\begin{eqnarray}
\label{MFeq2}
\{\widetilde{T_\alpha}  +  V_\alpha(x_\alpha) & + &
V_\alpha^{\mathrm{eff}}(y_\alpha) - E\} \widetilde{\psi_\alpha}(X_\alpha) =
  -  \frac{x_\alpha y_\alpha}{x_\beta y_\beta}V_\alpha^{(s)}(x_\alpha) \widetilde{\psi_\beta}(X_\beta) \nonumber \\
 & - & \left[ V_\beta^{(l)}(x_\beta) + V_3(x_3) - V_\alpha^{\mathrm{eff}}(y_\alpha) \right]\widetilde{\psi_\alpha}(X_\alpha),\ \ \alpha=1,2,
\end{eqnarray}
with $\widetilde{T_\alpha}=- \frac{\partial^2}{\partial y_\alpha^2} -
 \frac{\partial^2}{\partial x_\alpha^2}
-\left(\frac{1}{y_\alpha^2}+\frac{1}{x_\alpha^2}\right)
\frac{\partial}{\partial z_\alpha}(1-z_\alpha^2)\frac{\partial}{\partial z_\alpha}$.
The component $\widetilde{\psi_\alpha}$ satisfies zero boundary conditions on the lines $x_\alpha=0$, $y_\alpha=0$ and the asymptotic boundary condition obtained by multiplying formula~(\ref{zAs}) by $x_\alpha y_\alpha$ reads:
\begin{eqnarray}
\label{zAs2}
\widetilde{\psi_\alpha}(X_\alpha) & \sim & -\phi_{n_0\ell_0}(x_\alpha)
Y_{\ell_0 0}(\theta_\alpha,0)
 e^{-i\vartheta_{\ell_0}(y_\alpha,p_{n_0})} \delta_{\alpha,\alpha_0}\nonumber\\
& + & \sum\limits_{n\ell}
\phi_{n\ell}(x_\alpha)
Y_{\ell 0}(\theta_\alpha,0)
\sqrt{\frac{p_{n_0}}{p_n}}
S_{n\ell,n_0\ell_0}
e^{+i\vartheta_{\ell}(y_\alpha,p_n)}.
\end{eqnarray}
Another modification is done to make solutions of equations~(\ref{MFeq2}) real functions. We recast (\ref{zAs2}) into
 the form 
\begin{eqnarray}
\label{rAs}
\widetilde{\psi_\alpha}(X_\alpha) & \sim & -\phi_{n_0\ell_0}(x_\alpha)
Y_{\ell_0 0}(\theta_\alpha,0)
 \sin\left(\vartheta_{\ell_0}(y_\alpha,p_{n_0})\right) \delta_{\alpha,\alpha_0}\nonumber\\
& + & \sum\limits_{n\ell}
\phi_{n\ell}(x_\alpha)
Y_{\ell 0}(\theta_\alpha,0)
\sqrt{\frac{p_{n_0}}{p_n}}
K_{n\ell,n_0\ell_0}
\cos\left(\vartheta_{\ell}(y_\alpha,p_n)\right),
\end{eqnarray}
where the real numbers $K_{n\ell,n_0\ell_0}$ form the $K$-matrix.
One can show that a solution of the system of equations~(\ref{MFeq2}) with complex valued asymptotic boundary conditions~(\ref{zAs2})
is a linear combination of its solutions with boundary conditions~(\ref{rAs}) (standing waves) with different initial channels $\{\alpha_0,A_0\}$.
It gives the standard matrix relation between the $S$- and $K$-matrices
\begin{equation}
S = -(K+iI)^{-1}\cdot(K-iI).
\label{SK}
\end{equation}
Here $I$ is the identity matrix with linear size equal to the number of open channels $N_{\mathrm{ch}}$.

The boundary value problem~(\ref{MFeq2}),~(\ref{rAs}) is solved by the spline collocation method.
We solve equations in a box $[0,R_\alpha^x]\times[0,R_\alpha^y]\times[-1,1]$ for each component $\widetilde{\psi_\alpha}$.
As a basis set for expanding the component we use products of Hermite basis $S^3_5$ splines (splines of degree 5 with 2 continuous derivatives) in each coordinate.
Hermite basis splines form a basis in a space of $S^3_5$ splines on a given grid of knots. Each spline is a local function that is nonzero only on two adjoining intervals of the grid.
The boundary conditions~(\ref{rAs}) are implemented as follows.
For each open channel $\{\alpha_0,A_0\}$ we construct the driven system of equations by making substitutions
\begin{equation}
\widetilde{\psi_\alpha}(X_\alpha) = \widehat{\psi_\alpha}(X_\alpha) + \phi_{n_0\ell_0}(x_{\alpha})S(y_{\alpha})Y_{\ell_0 0}(\theta_\alpha,0)\delta_{\alpha, \alpha_0},
\end{equation}
where $S$ is the Hermite basis spline satisfying $S(R_{\alpha_0}^y)=1$:
\begin{eqnarray}
\label{MFfin}
\{\widetilde{T_\alpha} & + & V_\alpha(x_\alpha) +
V_\alpha^{\mathrm{eff}}(y_\alpha) - E\} \widehat{\psi_\alpha}(X_\alpha) =
  -  \frac{x_\alpha y_\alpha}{x_\beta y_\beta}V_\alpha^{(s)}(x_\alpha) \widehat{\psi_\beta}(X_\beta) \nonumber \\
 & - & \left[ V_\beta^{(l)}(x_\beta) + V_3(x_3) - V_\alpha^{\mathrm{eff}}(y_\alpha) \right]\widehat{\psi_\alpha}(X_\alpha) + f_\alpha(X_\alpha),\ \ \alpha=1,2.
\end{eqnarray}
Here the inhomogeneous term is given by
\begin{eqnarray}
f_\alpha(X_\alpha) = & - & \phi_{n_0\ell_0}(x_{\alpha_0}) Y_{\ell_0 0}(\theta_{\alpha_0},0) \Bigg\{ \left(-\frac{d^2}{d y_\alpha^2}+\frac{\ell_0(\ell_0+1)}{y_\alpha^2}+V_\beta^{(l)}+V_3-p_{n_0}^2\right) \delta_{\alpha,\alpha_0} \nonumber \\
& - & \frac{x_\alpha y_\alpha}{x_\beta y_\beta}V_\alpha^{(s)}(x_\alpha)\delta_{\beta,\alpha_0}\Bigg\}S(y_{\alpha_0}).
\end{eqnarray}
The system of equations~(\ref{MFfin}) is supplied  with zero boundary conditions $\widehat{\psi_\alpha}(X_\alpha)=0$ on the sides of the box $[0,R_\alpha^x]\times[0,R_\alpha^y]$.
The so obtained solution components $\widetilde{\psi_\alpha}^{(\alpha_0, A_0)}$, being a linear combination of physical solutions, behave asymptotically as
\begin{eqnarray}
\widetilde{\psi_\alpha}^{(\alpha_0, A_0)}(X_\alpha) & \sim & \sum\limits_{n\ell}
\phi_{n\ell}(x_\alpha)
Y_{\ell 0}(\theta_\alpha,0) \nonumber\\
& \times & \Big(
si_{n\ell,n_0\ell_0}
\sin\left(\vartheta_{\ell}(y_\alpha,p_n)\right)
+
co_{n\ell,n_0\ell_0}
\cos\left(\vartheta_{\ell}(y_\alpha,p_n)\right)
\Big),
\end{eqnarray}
with some coefficients $si_{n\ell,n_0\ell_0}$ and $co_{n\ell,n_0\ell_0}$.
These coefficients can be extracted by projecting the function $\widetilde{\psi_\alpha}^{(\alpha_0, A_0)}$ at some distant point $y_\alpha$ on two-body bound states
\begin{eqnarray}
\int_0^{R_\alpha^x}\mathrm{d}x_\alpha\int_{-1}^{1}\mathrm{d}\cos\theta_\alpha\,\phi_{n\ell}(x_\alpha)Y_{\ell 0}(\theta_\alpha,0)\widetilde{\psi_\alpha}^{(\alpha_0, A_0)}(X_\alpha)\nonumber\\
\qquad= si_{n\ell,n_0\ell_0}
\sin\left(\vartheta_{\ell}(y_\alpha,p_n)\right)
+
co_{n\ell,n_0\ell_0}
\cos\left(\vartheta_{\ell}(y_\alpha,p_n)\right)
\end{eqnarray}
and using extrapolation.
We obtain $N_{\mathrm{ch}}$ solutions $\widetilde{\psi_\alpha}^{(\alpha_0, A_0)}$ of driven equations corresponding to different open channels $\{\alpha_0,A_0\}$ for a given total energy $E$.
These solutions are linearly independent due to conditions $\widetilde{\psi_\alpha}^{(\alpha_0, A_0)}(X_\alpha)=\phi_{n_0\ell_0}(x_{\alpha})Y_{\ell 0}(\theta_\alpha,0)\delta_{\alpha, \alpha_0}$ at $X_\alpha=\{x_\alpha,R_\alpha^y,z_\alpha\}$.
The solution of interest $\widetilde{\psi_\alpha}$ is a linear combination of them.
The formula connecting $\widetilde{\psi_\alpha}$ and solutions $\widetilde{\psi_\alpha}^{(\alpha_0, A_0)}$ allows one to obtain all $K$-matrix elements from coefficients $si_{n\ell,n_0\ell_0}$ and $co_{n\ell,n_0\ell_0}$.

Equations~(\ref{MFfin}) are the final equations that are solved numerically.
The components are expanded in terms of products of Hermite basis splines:
\begin{equation}
\label{ser}
\widehat{\psi_\alpha}(X_\alpha) = \sum_{ijk}c_\alpha^{ijk}S_i(x_\alpha)S_j(y_\alpha)S_k(z_\alpha),\quad\alpha=1,2.
\end{equation}
Substituting the series~(\ref{ser}) into equations~(\ref{MFfin}) and requiring the resulting equalities to be satisfied in a number of points $\left(x_\alpha^\xi,\ y_\alpha^\eta,\ z_\alpha^\zeta\right)$ of a rectangular grid one gets the matrix equation on coefficients $c_\alpha^{ijk}$ of the form:
\hspace{-1cm}\begin{equation}
\label{mEq}
\left(
\begin{array}{cc}
[\bm{H}_1-E \bm{S}_1] & 0 \\
0 & [\bm{H}_2-E \bm{S}_2]
\end{array}
\right)
\left(
\begin{array}{c}
\bm{c_1} \\
\bm{c_2}
\end{array}
\right)
 =
\bm{R}
\left(
\begin{array}{c}
\bm{c_1} \\
\bm{c_2}
\end{array}
\right)
+
\left(
\begin{array}{c}
\bm{f_1} \\
\bm{f_2}
\end{array}
\right),
\end{equation}
where the matrices $\bm{H}_\alpha=\widetilde{\bm{T}_\alpha}+\bm{V}_\alpha+\bm{V}_\alpha^{\mathrm{eff}}$ and
\hspace{-2cm}\begin{equation}
\bm{R} =
-\left(
\begin{array}{cc}
0 & \bm{V}_1^{(s)} \\
\bm{V}_2^{(s)} & 0
\end{array}
\right)
-
\left(
\begin{array}{cc}
\bm{V}_2^{(l)}+\bm{V}_3-\bm{V}_1^{eff} & 0 \\
0 & \bm{V}_1^{(l)}+\bm{V}_3-\bm{V}_2^{eff}
\end{array}
\right)
\end{equation}
are discretized versions of the operators of the left- and right-hand sides of the system~(\ref{MFfin}).
Here for example the matrix $\bm{V}_\alpha$ has elements $V(x_\alpha^\xi)S_i(x_\alpha^\xi)S_j(y_\alpha^\eta)S_k(z_\alpha^\zeta)$ with rows enumerated by point numbers $\xi\eta\zeta$ and columns by basis function numbers $ijk$.
Right preconditioning by its left-hand side matrix turns the system of linear equations~(\ref{mEq}) into
\hspace{-4cm}{\begin{equation}
\label{SLAUfin}
\left\{
\bm{I}-\bm{R}
\left(
\begin{array}{cc}
[\bm{S}_1^{-1}\bm{H}_1-E \bm{I}]^{-1}\bm{S}_1^{-1} & 0 \\
0 & [\bm{S}_2^{-1}\bm{H}_2-E \bm{I}]^{-1}\bm{S}_2^{-1}
\end{array}
\right)
\right\}
\left(
\begin{array}{c}
\widetilde{\bm{c_1}} \\
\widetilde{\bm{c_2}}
\end{array}
\right)
=
\left(
\begin{array}{c}
\bm{f_1} \\
\bm{f_2}
\end{array}
\right).
\end{equation}}
This system of linear equations is solved by Arnoldi iterations in GMRES variant~\cite{saad03}.
On each iteration of the algorithm the most computationally expensive operation is a multiplication of the matrix of~(\ref{SLAUfin}) by a vector.
The matrix $\bm{R}$ is sparse, it has maximum of $2\times6\times6\times6=432$ nonzero elements in a row due to the locality property of Hermite basis splines.
To invert the matrix $[\bm{S}_\alpha^{-1}\bm{H}_\alpha-E\bm{I}]$ we use the algorithm which is known as ``tensor trick'' or matrix decomposition method (\cite{schelling89},\cite{bial93},\cite{roud00}). It provides a fast diagonalization of a matrix using its tensor product structure
\hspace{-3cm}\begin{equation}
\begin{array}{l}
[\bm{S}_\alpha^{-1}\bm{H}_\alpha-E\bm{I}]=(\bm{S}_{\alpha}^{x})^{-1}\bm{D}_{\alpha}^{x}\otimes \bm{I}\otimes \bm{I} + \bm{I}\otimes (\bm{S}_{\alpha}^{y})^{-1}\bm{D}_{\alpha}^{y}\otimes \bm{I} \\
+\Big(
(\bm{S}_{\alpha}^{x})^{-1}\bm{X}_{r2}\bm{S}_{\alpha}^{x}\otimes \bm{I}\otimes \bm{I} + \bm{I}\otimes (\bm{S}_{\alpha}^{y})^{-1}\bm{Y}_{r2}\bm{S}_{\alpha}^{y}\otimes \bm{I}
\Big)\cdot
(\bm{I}\otimes \bm{I}\otimes (\bm{S}_{\alpha}^{z})^{-1}\bm{D}_{\alpha}^{z})\\
+(\bm{S}_{\alpha}^{x})^{-1}\bm{V}_\alpha \bm{S}_{\alpha}^{x}\otimes \bm{I}\otimes \bm{I}+
\bm{I}\otimes (\bm{S}_{\alpha}^{y})^{-1}\bm{V}_\alpha^{eff}\bm{S}_{\alpha}^{y} \otimes \bm{I} - E\,\bm{I}\otimes \bm{I}\otimes \bm{I},
\end{array}
\end{equation}
where the matrices $\bm{S}^{x}_{\alpha}$, $\bm{D}^{x}_{\alpha}$, $\bm{S}^{y}_{\alpha}$, \ldots represent the ``one-dimensional'' matrices of basis splines and their second derivatives values at points of the grid in respective coordinates.
For example, $\bm{D}_\alpha^{x}$ is a matrix with elements $ S_i''(x_\alpha^\xi)$ enumerated by indices $i$ and $\xi$.
The matrices $\bm{V}_\alpha$, $\bm{X}_{r2}$, $\bm{Y}_{r2}$, \ldots are the diagonal matrices of pair and centrifugal potentials values.
For a more detailed description of a method we refer the reader to~\cite{roud00}.

\begin{figure}[t]

\center{\includegraphics[width=0.8\textwidth]{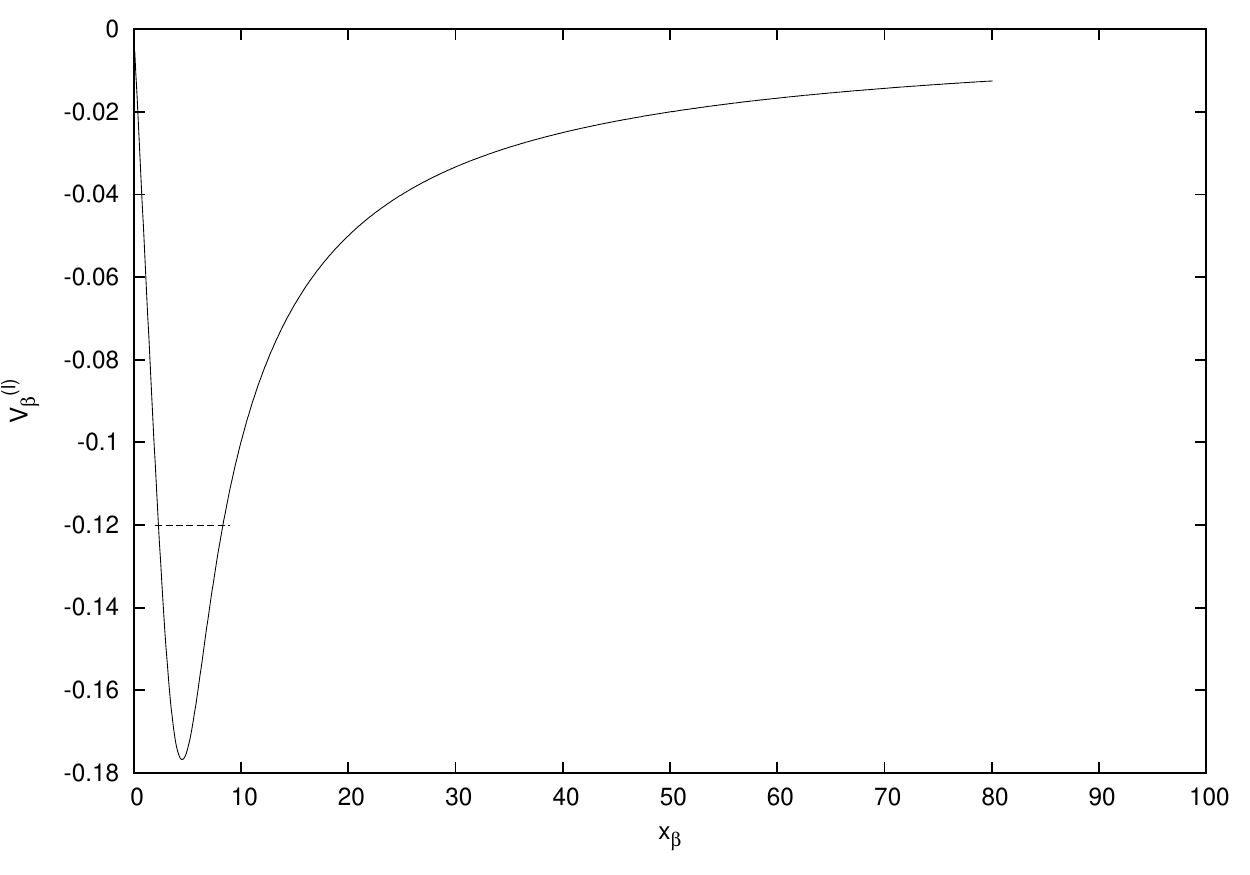}}

\caption{The long-range tail part of the potential $V_\beta=-1/x_\beta$ obtained by using the cut-off function~(\ref{Mcutoff}) with parameter $x_{0\beta}=3.0$. The dotted line denotes the ground state energy level of two particles interacting by this potential.}
\label{vl}

\end{figure}

At the end of this section we discuss the choice of the Merkuriev cut-off function~(\ref{Mcutoff}) parameter $x_{0\alpha}$.
We have found that for a given total energy $E$ there exists a lower bound for its values that has a simple physical interpretation.
Remember that the long-range part $V_\beta^{(l)}$ of the potential between particles of pair $\beta\ne\alpha$ is included in the left-hand side operator of equation $\alpha$ in~(\ref{MFeq}).
The potential $V_\beta^{(l)}$ is plotted in Figure~\ref{vl}, the ground state (g.s.) energy of two particles interacting via this potential is marked with a horizontal line.
The value of the g.s. energy is governed by the parameter $x_{0\beta}$.
The lower the parameter is, the deeper is the potential and the lower is the g.s. energy.
If the total energy of the three-body system $E$ is higher than this g.s. energy the operator in the left-hand side of equation $\alpha$ of~(\ref{MFeq})
\begin{equation}
T_\alpha + V_\alpha(x_\alpha) +
\sum_{\beta\ne\alpha}V_\beta^{(l)}(x_\beta)=
T_\beta +
\sum_{\beta\ne\alpha}V_\beta^{(l)}(x_\beta) + V_\alpha(x_\alpha)
\end{equation}
supports open channels in both arrangements $\alpha$ and $\beta$.
Subsequently, the outgoing waves corresponding to binary channels of pairing $\beta$ appear in the asymptote of $\psi_\alpha$.
It cancels one of the main advantages of Faddeev-Merkuriev equations approach --- the asymptotic uncoupling of components --- and should be avoided.
Thus, for a given total energy $E$ and a given $\beta$ the cut-off function parameter $x_{0\beta}$ should be bigger than some critical value $x_{c}$ that is defined as follows:
the ground state energy of a two-body Hamiltonian with the potential $V^{(l)}_\beta$ for $x_{0\beta}=x_{c}$ should be equal to $E$.

There is no upper bound on the values of $x_{0\beta}$, but we observed that the bigger is the value of the parameter, the bigger basis in angular coordinate $z_\alpha$ sizes are to be taken to obtain the desired accuracy.
Summarizing, we propose the following algorithm of choosing the cut-off function parameter $x_{0\beta}$:
\begin{enumerate}
\item
\label{vvv}
For a total energy interval $[E_{\min},E_{\max}]$ of interest find the critical values of parameters $x_{0\beta}$ for the total energy value $E_{\max}$ for each $\beta$. The lowest energy levels of the two-body Hamiltonian with the potential $V_\beta^{(l)}$ can be calculated by any simple two-body code.
\item Take the value of the parameter $x_{0\beta}$ only slightly bigger than the critical value calculated at step~(\ref{vvv}), since bigger values of $x_{0\beta}$ require employing larger grids.
\end{enumerate}

\section{Results}
\subsection{$e^-e^+\bar{p}$ and $e^+e^-p$ scattering}
Positron--hydrogen atom scattering is the simplest example of positron-atom scattering process.
Many calculations are presented in the literature, among them are detailed calculations using the FM equations for both low energies~\cite{lazau18, hu99,yak07} and wide energy region including that above the ionisation threshold~\cite{lazau17}.
Other methods include hyperspherical calculations~\cite{igara94,ward12}, variational~\cite{humber97,gien99}, close coupling calculations using two center basis functions expansion~\cite{mitroy95,kadyr02}.
The renewed interest in studying the reactions involving positron, electron and (anti)proton is motivated by experiments on antimatter that take place at CERN~\cite{lazau18}.
The key role in antimatter formation plays the reaction of antihydrogen formation via antiproton ($\bar{p}$) impact of positronium (Ps, the bound state of $e^+$ and $e^-$) atom.
Due to the symmetry in particle charges, the cross sections in $e^+e^-p$ and $e^-e^+\bar{p}$ systems are identical.
Further in this section we refer to the $e^+e^-\bar{p}$ system.

We have calculated K-matrix elements of all possible scattering processes in $e^+e^-\bar{p}$ system in the total energy range from $-0.49973\ a.u.$ to  $-0.05553\ a.u.$ with the energy step of calculation $0.0007\ a.u.$
In this interval elastic, excitations and rearrangement processes leading to $\bar{\mbox{H}}$($n=1,2$) and Ps($n=1,2$) atom states are possible.
The maximum linear size of K-matrix equals to 6.
We proceed as follows: the whole energy region is divided into subintervals associated with energy thresholds.
For every interval we choose appropriate discretisation and cut-off function parameters according to the rules described in the previous section.
This is done to minimize the computational cost, since generally  higher energy calculations require
\begin{itemize}
\item bigger box sizes in $x_\alpha$, $y_\alpha$ and more basis functions involved
\item bigger cut-off parameters $x_{0\alpha}$, which in turn increases the number of functions in the angular coordinate basis.
\end{itemize}
The energy intervals and corresponding discretisation and cut-off function parameters are given in table~\ref{H_discr}.
In this table and in further applications we use shortcuts $\bar{\mbox{H}}(n)$ and $\bar{\mbox{H}}(n, \ell)$
for the atom states with principal quantum number $n$ and angular momentum $\ell$.
\begin{table}[t!]
\centering
\begin{tabular}{c|c|c|c}
\hline
 & $\bar{\mbox{H}}$(1)--Ps(1) & Ps(1)--$\bar{\mbox{H}}$(2) & $\bar{\mbox{H}}$(2)--$\bar{\mbox{H}}$(3) \\
\hline
First component\\
\hline
$x_{\max}$ & 17.7 & 28.3 & 70.7 \\
$y_{\max}$ & 42.4 & 84.9 & 141 \\
$n_x$ & 45 & 60 & 75 \\
$n_y$ & 120 & 300 & 510 \\
$n_z$ & 24 & 24 & 24 \\
$x_0$ & 1.4 & 3.5 & 8.5 \\
\hline
Second component\\
\hline
$x_{\max}$ & 35 & 70  & 100 \\
$y_{\max}$ & 17.5 & 35 & 115 \\
$n_x$ & 60 & 75 & 75 \\
$n_y$ & 60 & 105  & 600 \\
$n_z$ & 18 & 30  & 45 \\
$x_0$ & 1 & 4 & 8 \\
\end{tabular}
\caption{Discretisation and cut-off function parameters used in calculations of $e^+e^-\bar{p}$ system cross sections. The column headings denote the energy intervals associated with atoms excitations thresholds.}
\label{H_discr}
\end{table}

In the calculations we hold the accuracy level such that the maximum error does not exceed 1\%.
Tables~\ref{conv_H} and~\ref{conv_H_2} illustrate the convergence of different cross sections at energy $E=-0.0572\  a.u.$ with respect to discretisation and cut-off function parameters.
To control our calculations, we have also used the value
\[
\mathfrak{K}=||K-K^T||/||K||
\]
with the Frobenius norm of a matrix, as a measure of K-matrix asymmetry.
The values of $\mathfrak{K}$ are shown in tables~\ref{conv_H} and~\ref{conv_H_2}.
\begin{landscape}
\begin{table}
{\tiny
\begin{tabular}{c|c|c|c|c|c|c|c|c|c|c|c|c}
 & $\sigma_{\bar{\mbox{H}}(1)- \bar{\mbox{H}}(1)}$ & $\sigma_{\bar{\mbox{H}}(1)- \bar{\mbox{H}}(2,s)}$ & $\sigma_{\bar{\mbox{H}}(1)- \bar{\mbox{H}}(2,p)}$ & $\sigma_{\bar{\mbox{H}}(1)- \mbox{Ps}(1)}$ & $\sigma_{\bar{\mbox{H}}(2,s)- \bar{\mbox{H}}(1)}$ & $\sigma_{\bar{\mbox{H}}(2,s)- \mbox{Ps}(1)}$ & $\sigma_{\mbox{Ps}(1)- \bar{\mbox{H}}(1)}$ & $\sigma_{\mbox{Ps}(1)- \bar{\mbox{H}}(2)}$ & $\sigma_{\mbox{Ps}(1)- \mbox{Ps}(1)}$ & $\sigma_{\mbox{Ps}(2)- \mbox{Ps}(2)}$ & $\sigma_{\mbox{Ps}(2)- \bar{\mbox{H}}(n\le2)}$ & $\mathfrak{K}$ \\
\hline
$x_{max}$ & & & & & & & & & & & & \\
70.0 & 0.11392 & 0.0023206 & 0.0018759 & 0.005665 & 0.015184 & 0.37833 & 0.0064832 & 0.10521 & 4.8348 & 270.21 & 16.104 & 0.42704 \\
80.0 & 0.11392 & 0.0023966 & 0.0019211 & 0.0056769 & 0.015648 & 0.39138 & 0.0065249 & 0.1099 & 4.833 & 194.17 & 12.235 & 0.24687 \\
90.0 & 0.11392 & 0.0024092 & 0.0019205 & 0.0056744 & 0.015725 & 0.39225 & 0.0065278 & 0.11046 & 4.8326 & 184.23 & 12.122 & 0.11033 \\
100.0 & 0.11392 & 0.0024076 & 0.0019189 & 0.0056738 & 0.015714 & 0.39165 & 0.0065281 & 0.11034 & 4.8327 & 183.38 & 12.108 & 0.096057 \\
110.0 & 0.11392 & 0.0024063 & 0.0019188 & 0.005674 & 0.015705 & 0.39142 & 0.0065287 & 0.11029 & 4.8329 & 183.81 & 12.105 & 0.10578 \\
$y_{max}$ & & & & & & & & & & & & \\
170.0 & 0.11402 & 0.0023717 & 0.0019656 & 0.0056777 & 0.015481 & 0.39408 & 0.0065316 & 0.11075 & 4.832 & 183.37 & 12.052 & 0.098401 \\
180.0 & 0.114 & 0.0023633 & 0.0019241 & 0.0056804 & 0.015461 & 0.39297 & 0.0065341 & 0.11053 & 4.8324 & 183.37 & 12.091 & 0.097098 \\
190.0 & 0.11395 & 0.0023846 & 0.0019089 & 0.0056846 & 0.015583 & 0.39205 & 0.0065375 & 0.11037 & 4.8327 & 183.39 & 12.095 & 0.096431 \\
200.0 & 0.11392 & 0.0024076 & 0.0019189 & 0.0056738 & 0.015714 & 0.39165 & 0.0065281 & 0.11034 & 4.8327 & 183.38 & 12.108 & 0.096057 \\
210.0 & 0.11385 & 0.0023946 & 0.0018732 & 0.0056825 & 0.01568 & 0.3918 & 0.006535 & 0.11042 & 4.8325 & 183.37 & 12.124 & 0.095862 \\
$n_{x}$ & & & & & & & & & & & & \\
30 & 0.11315 & 0.0024155 & 0.0019283 & 0.0057331 & 0.015828 & 0.39656 & 0.0066087 & 0.11167 & 4.8343 & 186.54 & 11.755 & 0.10776 \\
45 & 0.11385 & 0.002411 & 0.0019194 & 0.0056766 & 0.015732 & 0.39186 & 0.0065335 & 0.11038 & 4.8317 & 183.67 & 12.078 & 0.097658 \\
60 & 0.11391 & 0.0024083 & 0.0019188 & 0.0056751 & 0.015717 & 0.39176 & 0.0065298 & 0.11037 & 4.8326 & 183.41 & 12.105 & 0.096165 \\
75 & 0.11392 & 0.0024076 & 0.0019189 & 0.0056738 & 0.015714 & 0.39165 & 0.0065281 & 0.11034 & 4.8327 & 183.38 & 12.108 & 0.096057 \\
90 & 0.11392 & 0.0024076 & 0.0019189 & 0.0056736 & 0.015714 & 0.39162 & 0.0065282 & 0.11035 & 4.8328 & 183.38 & 12.109 & 0.096093 \\
$n_{y}$ & & & & & & & & & & & & \\
420 & 0.11354 & 0.0024063 & 0.0019183 & 0.0056739 & 0.015703 & 0.39167 & 0.0065284 & 0.11036 & 4.8327 & 183.38 & 12.11 & 0.096054 \\
450 & 0.11373 & 0.002407 & 0.0019188 & 0.0056739 & 0.015708 & 0.39166 & 0.0065284 & 0.11035 & 4.8327 & 183.38 & 12.11 & 0.096055 \\
480 & 0.11384 & 0.0024074 & 0.0019189 & 0.005674 & 0.015712 & 0.39165 & 0.0065284 & 0.11035 & 4.8327 & 183.38 & 12.109 & 0.096056 \\
510 & 0.11392 & 0.0024076 & 0.0019189 & 0.0056738 & 0.015714 & 0.39165 & 0.0065281 & 0.11034 & 4.8327 & 183.38 & 12.108 & 0.096057 \\
540 & 0.11397 & 0.0024078 & 0.0019188 & 0.0056739 & 0.015716 & 0.39164 & 0.006528 & 0.11034 & 4.8327 & 183.38 & 12.108 & 0.096058 \\
$n_{z}$ & & & & & & & & & & & & \\
15 & 0.11405 & 0.0024317 & 0.001936 & 0.005723 & 0.015875 & 0.41145 & 0.0066155 & 0.11639 & 4.7913 & 211.2 & 13.616 & 0.35515 \\
18 & 0.11398 & 0.0024236 & 0.0019208 & 0.0056932 & 0.015823 & 0.40079 & 0.006563 & 0.11321 & 4.8127 & 194.27 & 12.942 & 0.26182 \\
21 & 0.11394 & 0.002413 & 0.0019171 & 0.00568 & 0.01575 & 0.39465 & 0.0065396 & 0.11132 & 4.8252 & 186.65 & 12.389 & 0.17454 \\
24 & 0.11392 & 0.0024076 & 0.0019189 & 0.0056738 & 0.015714 & 0.39165 & 0.0065281 & 0.11034 & 4.8327 & 183.38 & 12.108 & 0.096057 \\
27 & 0.1139 & 0.0024067 & 0.0019217 & 0.0056701 & 0.015709 & 0.39027 & 0.0065214 & 0.10987 & 4.8375 & 182 & 12.005 & 0.049207 \\
$x_{0}$ & & & & & & & & & & & & \\
8.0 & 0.11404 & 0.0033197 & 0.0076791 & 0.0046552 & 0.0042209 & 0.32523 & 0.0073201 & 0.30122 & 5.0731 & 444.54 & 23.429 & 1.3065 \\
8.5 & 0.11377 & 0.0024429 & 0.0026296 & 0.0058196 & 0.016796 & 0.44711 & 0.006609 & 0.1328 & 4.8479 & 199.52 & 14.579 & 0.65552 \\
9.0 & 0.11394 & 0.0024398 & 0.0017586 & 0.0056412 & 0.015703 & 0.37821 & 0.0064939 & 0.10462 & 4.8426 & 177.33 & 11.52 & 0.19912 \\
9.5 & 0.11395 & 0.0023476 & 0.0020437 & 0.0056189 & 0.014731 & 0.38564 & 0.0065479 & 0.11389 & 4.85 & 183.59 & 12.32 & 0.13869 \\
10.0 & 0.11395 & 0.0024016 & 0.0018837 & 0.0056341 & 0.015287 & 0.38313 & 0.0065232 & 0.10921 & 4.8434 & 180.02 & 11.895 & 0.041614 \\
10.5 & 0.11391 & 0.0024128 & 0.0019318 & 0.0056781 & 0.015796 & 0.39302 & 0.0065243 & 0.11053 & 4.8388 & 180.75 & 12.068 & 0.0089676 \\
11.0 & 0.11392 & 0.0024119 & 0.0019273 & 0.0056761 & 0.015759 & 0.39263 & 0.0065264 & 0.11051 & 4.837 & 180.45 & 12.042 & 0.011675 \\
11.5 & 0.11392 & 0.0024124 & 0.0019288 & 0.0056777 & 0.015759 & 0.39309 & 0.0065294 & 0.11064 & 4.8349 & 179.8 & 12.045 & 0.048489 \\
12.0 & 0.11392 & 0.0024076 & 0.0019189 & 0.0056738 & 0.015714 & 0.39165 & 0.0065281 & 0.11034 & 4.8327 & 183.38 & 12.108 & 0.096057 \\
12.5 & 0.11392 & 0.0024087 & 0.0019224 & 0.0056761 & 0.015725 & 0.39223 & 0.0065314 & 0.11048 & 4.8302 & 181.47 & 12.025 & 0.034421 \\
13.0 & 0.11392 & 0.0024085 & 0.0019231 & 0.0056768 & 0.015725 & 0.39229 & 0.0065336 & 0.11051 & 4.8274 & 181.18 & 12.007 & 0.023126 \\
13.5 & 0.11392 & 0.0024083 & 0.0019235 & 0.0056772 & 0.015723 & 0.39225 & 0.0065356 & 0.1105 & 4.8244 & 181.05 & 11.993 & 0.018097 \\
14.0 & 0.11392 & 0.002408 & 0.0019238 & 0.0056774 & 0.015721 & 0.39215 & 0.0065375 & 0.11049 & 4.821 & 180.98 & 11.981 & 0.0151 \\
\end{tabular}
}
\caption{Convergence of cross sections for scattering in $e^+e^-\bar{\mbox{p}}$ system with respect to discretization and cut-off function parameters related to the first component $\psi_1$.}
\label{conv_H}
\end{table}
\begin{table}
{\tiny
\begin{tabular}{c|c|c|c|c|c|c|c|c|c|c|c|c}
 & $\sigma_{\bar{\mbox{H}}(1)- \bar{\mbox{H}}(1)}$ & $\sigma_{\bar{\mbox{H}}(1)- \bar{\mbox{H}}(2,s)}$ & $\sigma_{\bar{\mbox{H}}(1)- \bar{\mbox{H}}(2,p)}$ & $\sigma_{\bar{\mbox{H}}(1)- \mbox{Ps}(1)}$ & $\sigma_{\bar{\mbox{H}}(2,s)- \bar{\mbox{H}}(1)}$ & $\sigma_{\bar{\mbox{H}}(2,s)- \mbox{Ps}(1)}$ & $\sigma_{\mbox{Ps}(1)- \bar{\mbox{H}}(1)}$ & $\sigma_{\mbox{Ps}(1)- \bar{\mbox{H}}(2)}$ & $\sigma_{\mbox{Ps}(1)- \mbox{Ps}(1)}$ & $\sigma_{\mbox{Ps}(2)- \mbox{Ps}(2)}$ & $\sigma_{\mbox{Ps}(2)- \bar{\mbox{H}}(n\le2)}$ & $\mathfrak{K}$ \\
\hline
\hline
$x_{max}$ & & & & & & & & & & & & \\
70.0 & 0.11394 & 0.0023579 & 0.0019599 & 0.0056999 & 0.015557 & 0.38978 & 0.006538 & 0.11023 & 4.8343 & 183.46 & 11.705 & 0.088877 \\
80.0 & 0.11395 & 0.0024123 & 0.0019042 & 0.0056629 & 0.015677 & 0.39072 & 0.0065238 & 0.10995 & 4.8331 & 183.45 & 11.986 & 0.093915 \\
90.0 & 0.11392 & 0.0024073 & 0.0019258 & 0.0056769 & 0.015732 & 0.39213 & 0.0065278 & 0.11047 & 4.8327 & 183.48 & 12.113 & 0.09549 \\
100.0 & 0.11392 & 0.0024076 & 0.0019189 & 0.0056738 & 0.015714 & 0.39165 & 0.0065281 & 0.11034 & 4.8327 & 183.38 & 12.108 & 0.096057 \\
110.0 & 0.11391 & 0.002409 & 0.00192 & 0.0056742 & 0.015723 & 0.39184 & 0.0065281 & 0.11038 & 4.8327 & 183.41 & 12.117 & 0.096168 \\
$y_{max}$ & & & & & & & & & & & & \\
90.0 & 0.11389 & 0.00247 & 0.0020719 & 0.0057501 & 0.014244 & 0.36754 & 0.007186 & 0.16077 & 4.844 & 186.28 & 8.731 & 0.5278 \\
100.0 & 0.11387 & 0.002293 & 0.0020065 & 0.005741 & 0.015615 & 0.39253 & 0.006386 & 0.10129 & 4.8324 & 184.87 & 8.8342 & 0.35352 \\
110.0 & 0.11395 & 0.0024066 & 0.0018722 & 0.0056544 & 0.015669 & 0.38917 & 0.0064704 & 0.10832 & 4.8314 & 181.46 & 12.038 & 0.18133 \\
120.0 & 0.11393 & 0.0023884 & 0.001923 & 0.00567 & 0.015393 & 0.38183 & 0.0065911 & 0.1099 & 4.8393 & 183.51 & 13.255 & 0.16197 \\
130.0 & 0.11392 & 0.0023832 & 0.0019517 & 0.0056963 & 0.015605 & 0.39306 & 0.0065423 & 0.10991 & 4.8353 & 186.18 & 11.365 & 0.076749 \\
140.0 & 0.11394 & 0.0024358 & 0.0019393 & 0.0056783 & 0.015893 & 0.39998 & 0.0065331 & 0.1127 & 4.8298 & 184.25 & 11.062 & 0.083529 \\
150.0 & 0.11393 & 0.0024153 & 0.0019091 & 0.0056699 & 0.015761 & 0.38884 & 0.0065236 & 0.10948 & 4.8343 & 181.65 & 12.537 & 0.094786 \\
160.0 & 0.11391 & 0.0023749 & 0.0019151 & 0.0056802 & 0.0155 & 0.38659 & 0.0065357 & 0.10871 & 4.8354 & 182.8 & 12.705 & 0.101 \\
170.0 & 0.11391 & 0.0023921 & 0.0019395 & 0.0056847 & 0.015616 & 0.38949 & 0.0065397 & 0.10934 & 4.8356 & 185.15 & 11.564 & 0.095714 \\
180.0 & 0.11394 & 0.002419 & 0.0019339 & 0.0056795 & 0.015789 & 0.39555 & 0.0065347 & 0.11131 & 4.8325 & 184.81 & 11.401 & 0.092771 \\
190.0 & 0.11395 & 0.0024091 & 0.0019131 & 0.0056671 & 0.015721 & 0.39086 & 0.0065199 & 0.11024 & 4.832 & 182.97 & 12.303 & 0.094655 \\
200.0 & 0.11393 & 0.0023862 & 0.001916 & 0.0056784 & 0.015571 & 0.38777 & 0.0065339 & 0.109 & 4.8354 & 182.79 & 12.515 & 0.097142 \\
210.0 & 0.11391 & 0.0023951 & 0.0019344 & 0.0056847 & 0.015632 & 0.39134 & 0.0065404 & 0.10999 & 4.8344 & 183.84 & 11.764 & 0.097132 \\
220.0 & 0.11391 & 0.0024129 & 0.0019341 & 0.0056815 & 0.01575 & 0.39237 & 0.0065364 & 0.11027 & 4.8342 & 184.12 & 11.563 & 0.096512 \\
230.0 & 0.11392 & 0.0024076 & 0.0019189 & 0.0056738 & 0.015714 & 0.39165 & 0.0065281 & 0.11034 & 4.8327 & 183.38 & 12.108 & 0.096057 \\
240.0 & 0.11393 & 0.0023906 & 0.0019161 & 0.0056734 & 0.015601 & 0.38804 & 0.0065273 & 0.10922 & 4.8342 & 182.89 & 12.398 & 0.095718 \\
$n_{x}$ & & & & & & & & & & & & \\
30 & 0.1156 & 0.00227 & 0.0018007 & 0.0056725 & 0.014761 & 0.39132 & 0.0065013 & 0.11016 & 4.833 & 183.34 & 12.072 & 0.095884 \\
45 & 0.11495 & 0.0024034 & 0.0019084 & 0.0056662 & 0.015692 & 0.39168 & 0.0065242 & 0.11034 & 4.8328 & 183.39 & 12.107 & 0.095996 \\
60 & 0.11395 & 0.0024074 & 0.0019179 & 0.0056726 & 0.015713 & 0.39165 & 0.0065278 & 0.11034 & 4.8327 & 183.38 & 12.108 & 0.096045 \\
75 & 0.11392 & 0.0024076 & 0.0019189 & 0.0056738 & 0.015714 & 0.39165 & 0.0065281 & 0.11034 & 4.8327 & 183.38 & 12.108 & 0.096057 \\
90 & 0.11393 & 0.0024077 & 0.0019185 & 0.0056724 & 0.015712 & 0.39164 & 0.0065263 & 0.11034 & 4.8327 & 183.38 & 12.108 & 0.09606 \\
$n_{y}$ & & & & & & & & & & & & \\
510 & 0.11399 & 0.0024076 & 0.001919 & 0.0056694 & 0.015714 & 0.39182 & 0.0065224 & 0.11042 & 4.8326 & 183.4 & 12.087 & 0.096081 \\
540 & 0.11399 & 0.0024072 & 0.0019188 & 0.0056685 & 0.015711 & 0.39175 & 0.0065215 & 0.11039 & 4.8327 & 183.4 & 12.095 & 0.09607 \\
570 & 0.11394 & 0.0024076 & 0.0019189 & 0.0056723 & 0.015714 & 0.3917 & 0.0065263 & 0.11037 & 4.8327 & 183.39 & 12.102 & 0.096062 \\
600 & 0.11392 & 0.0024076 & 0.0019189 & 0.0056738 & 0.015714 & 0.39165 & 0.0065281 & 0.11034 & 4.8327 & 183.38 & 12.108 & 0.096057 \\
630 & 0.11392 & 0.0024075 & 0.0019188 & 0.0056739 & 0.015713 & 0.3916 & 0.0065282 & 0.11032 & 4.8328 & 183.37 & 12.114 & 0.096053 \\
$n_{z}$ & & & & & & & & & & & & \\
36 & 0.11441 & 0.002408 & 0.0019187 & 0.005678 & 0.015717 & 0.39161 & 0.0065335 & 0.11035 & 4.8327 & 183.38 & 12.111 & 0.09623 \\
39 & 0.11418 & 0.0024079 & 0.0019188 & 0.0056762 & 0.015716 & 0.39162 & 0.0065312 & 0.11035 & 4.8327 & 183.38 & 12.109 & 0.096127 \\
42 & 0.11404 & 0.0024077 & 0.0019188 & 0.0056748 & 0.015715 & 0.39164 & 0.0065295 & 0.11034 & 4.8327 & 183.38 & 12.109 & 0.096099 \\
45 & 0.11392 & 0.0024076 & 0.0019189 & 0.0056738 & 0.015714 & 0.39165 & 0.0065281 & 0.11034 & 4.8327 & 183.38 & 12.108 & 0.096057 \\
48 & 0.11382 & 0.0024075 & 0.0019189 & 0.0056729 & 0.015714 & 0.39165 & 0.006527 & 0.11034 & 4.8327 & 183.38 & 12.108 & 0.096018 \\
51 & 0.11374 & 0.0024075 & 0.001919 & 0.0056723 & 0.015713 & 0.39166 & 0.0065261 & 0.11034 & 4.8328 & 183.38 & 12.107 & 0.095978 \\
54 & 0.11368 & 0.0024074 & 0.001919 & 0.0056717 & 0.015713 & 0.39166 & 0.0065253 & 0.11034 & 4.8328 & 183.38 & 12.106 & 0.095942 \\
57 & 0.11419 & 0.0024067 & 0.0019185 & 0.0056749 & 0.015709 & 0.39158 & 0.006529 & 0.11033 & 4.8327 & 183.38 & 12.107 & 0.096171 \\
$x_{0}$ & & & & & & & & & & & & \\
6.0 & 0.11314 & 0.0025699 & 0.0024345 & 0.0059803 & 0.014396 & 0.38186 & 0.0069249 & 0.14078 & 4.8302 & 13263 & 907.81 & 0.37535 \\
6.5 & 0.11339 & 0.0025983 & 0.0021622 & 0.0057896 & 0.016454 & 0.4292 & 0.006689 & 0.12736 & 4.8258 & 228.04 & 5.2034 & 0.18133 \\
7.0 & 0.11358 & 0.0024108 & 0.0019345 & 0.0056821 & 0.015736 & 0.39353 & 0.0065382 & 0.11078 & 4.8332 & 179.76 & 11.774 & 0.086407 \\
7.5 & 0.11373 & 0.0024112 & 0.0019325 & 0.0056808 & 0.015745 & 0.39351 & 0.0065356 & 0.11073 & 4.833 & 179.59 & 11.984 & 0.10141 \\
8.0 & 0.11392 & 0.0024076 & 0.0019189 & 0.0056738 & 0.015714 & 0.39165 & 0.0065281 & 0.11034 & 4.8327 & 183.38 & 12.108 & 0.096057 \\
8.5 & 0.11414 & 0.0024084 & 0.0019229 & 0.0056756 & 0.015723 & 0.39229 & 0.0065296 & 0.11047 & 4.8327 & 180.96 & 12.016 & 0.013574 \\
9.0 & 0.1144 & 0.0024081 & 0.0019231 & 0.0056754 & 0.015722 & 0.39238 & 0.0065291 & 0.11049 & 4.8327 & 180.72 & 12.023 & 0.0042565 \\
9.5 & 0.11471 & 0.0024078 & 0.0019229 & 0.0056749 & 0.01572 & 0.39239 & 0.0065285 & 0.1105 & 4.8326 & 180.65 & 12.029 & 0.0018316 \\
10.0 & 0.11506 & 0.0024074 & 0.0019226 & 0.0056744 & 0.015718 & 0.39239 & 0.0065278 & 0.1105 & 4.8326 & 180.62 & 12.034 & 0.0011176 \\
\end{tabular}
}
\caption{Convergence of cross sections for scattering in $e^+e^-\bar{\mbox{H}}$ system with respect to discretization and cut-off function parameters related to the second component $\psi_2$.}
\label{conv_H_2}
\end{table}
\end{landscape}

In table~\ref{others} we compare our results with tabulated values of other authors and give the values of some additional cross sections for further references.
The agreement between our results and that of other calculations is quite good.
The only noticeable  disagreement is in some values of cross sections for $\bar{\mbox{H}}(1)$ formation, our results being a few percent lower than that of other authors.
We note that  the calculation of $\bar{\mbox{H}}(1)$ formation amplitudes should be done with extra accuracy otherwise
it may be prone to errors.
The reason is that the wave function of $\bar{\mbox{H}}(1)$ formation process is highly oscillatory in the asymptotic region of positron flying off the ground state of antihydrogen due to the large positron momentum $p_{1}=\sqrt{E-\varepsilon_{\bar{\mbox{{\tiny H}}}(1)}}$ in~(\ref{asympt3D}).
Therefore the extra precision of numerical solution is required to guaranty the extraction of the amplitude with controlled accuracy.
Our results are 
converged with respect to all numerical solution parameters, which we additionally checked for $\bar{\mbox{H}}(1)$ formation cross sections. 

\begin{table}[t!]
{\small
\begin{tabular}{c|c|c|c|c|c|c|c|c}
\backslashbox{}{$E$, a.u.} & 0.27026 & 0.28140 & 0.32017 & 0.36145 & 0.385 & 0.40 & 0.415 & 0.42\\
\hline
$\sigma_{\bar{\mbox{H}}(1)\rightarrow \bar{\mbox{H}}(1)}$ & 0.0353 & 0.0417 & 0.0634 & 0.0836 & 0.0944 & 0.100 & 0.105 & 0.107\\
\cite{mitroy95} & & & 0.0651 & 0.0844 & & 0.100 &  \\
\cite{hu99} & 0.0372 & 0.0429 & 0.0649 & 0.0866 & 0.090 & 0.096 & 0.099 & 0.101\\
\cite{gien99} & & 0.0431 & 0.0650 & 0.0856 & \\
\hline
$\sigma_{\bar{\mbox{H}}(1)\rightarrow Ps(1)}$ & 0.00412 & 0.00430 & 0.00487 & 0.00562 & 0.00565 & 0.00572 & 0.00575 & 0.00574\\
\cite{mitroy95} & & & 0.00490 & 0.00567 & & 0.00581 & \\
\cite{hu99} & 0.00410 & 0.00439 & 0.00487 & 0.00557 & \\
\cite{gien99} & & 0.00422 & 0.00481 & 0.00554 & \\
\hline
$\sigma_{Ps(1)\rightarrow Ps(1)}$ & 3.49 & 7.06 & 9.87 & 8.31 & 7.11 & 6.44 & 5.82 & 5.62 \\
\cite{mitroy95} & & & 9.87 & 8.32 & & 6.45 & \\
\cite{hu99} & 3.500 & 7.060 & 9.866 & 8.312 & 7.09 & 6.44 & 5.83 & 5.63\\
\cite{gien99} & & 6.936 & 9.868 & 8.332 &  \\
\hline
$\sigma_{Ps(1)\rightarrow \bar{\mbox{H}}(1)}$ & 0.0272 & 0.0191 & 0.0111 & 0.0091 & 0.00806 & 0.00763 & 0.00724 & 0.00709 \\
\cite{hu99} & 0.0274 & 0.0195 & 0.0111 & 0.0091 & 0.00815 & 0.00780 & 0.00729 & 0.00715\\
\hline
$\sigma_{\bar{\mbox{H}}(1)\rightarrow \bar{\mbox{H}}(2, s)}$ & & & & & 0.000662 & 0.00137 & 0.00206 & 0.00228\\
\hline
$\sigma_{\bar{\mbox{H}}(1)\rightarrow \bar{\mbox{H}}(2, p)}$ & & & & & 0.000399 & 0.000236 & 0.000421 & 0.000582 \\
\hline
$\sigma_{\bar{\mbox{H}}(2,s)\rightarrow Ps(1, s)}$ & & & & & 1.26 & 0.576 & 0.477 & 0.475 \\
\hline
$\sigma_{\bar{\mbox{H}}(2,s)\rightarrow \bar{\mbox{H}}(1, s)}$ & & & & & 0.0249 & 0.0217 & 0.0212 & 0.0212 \\
\hline
$\sigma_{Ps(1)\rightarrow \bar{\mbox{H}}(2, s)}$ & & & & & 0.0476 & 0.0484 & 0.0581 & 0.0631 \\
\hline
$\sigma_{Ps(1)\rightarrow \bar{\mbox{H}}(2, p)}$ & & & & & 0.0390 & 0.0484 & 0.0512 & 0.0519 \\
\hline
\end{tabular}
}
\caption{Scattering cross sections in $e^-e^+\bar{p}$ system, present results and that of other authors (the energy is measured from the -0.49973 a.u. $\bar{\mbox{H}}$(1) threshold)}
\label{others}
\end{table}

\begin{figure}[t]
\center{\includegraphics[width=0.99\textwidth]{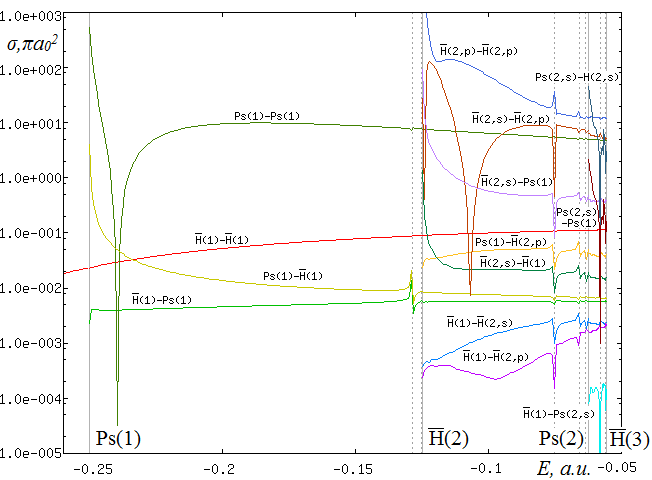}}
\caption{Cross sections in $e^-e^+\bar{p}$ system. Vertical solid lines denote binary thresholds, vertical dashed lines mark resonances positions.}
\label{CS_H}

\end{figure}

Cross sections for the processes with positron or antiproton colliding with antihydrogen or positronium respectively are presented in Fig.~\ref{CS_H}.
We see that some cross sections associated with excited antihydrogen or positronium states are large, for example the $\bar{\mbox{H}}(2,s)$ - Ps(1) and Ps$(2, s)$ - $\bar{\mbox{H}}(2, s)$ cross sections for rearrangement processes of positronium and antihydrogen formation.
The positronium ground state elastic channel cross section is also large and is almost constant along a wide energy interval.

\begin{figure}[t]
\begin{minipage}[h]{0.32\linewidth}
\center{\includegraphics[width=0.99\textwidth]{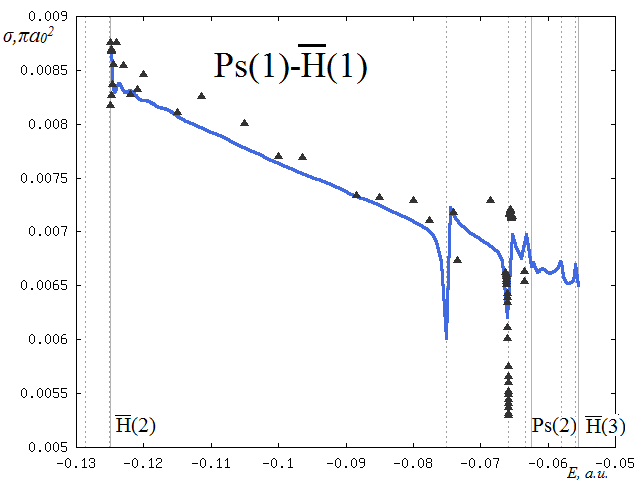}}\\
Ps(1)--$\bar{\mbox{H}}(1)$
\end{minipage}
\hfill
\begin{minipage}[h]{0.32\linewidth}
\center{\includegraphics[width=0.99\textwidth]{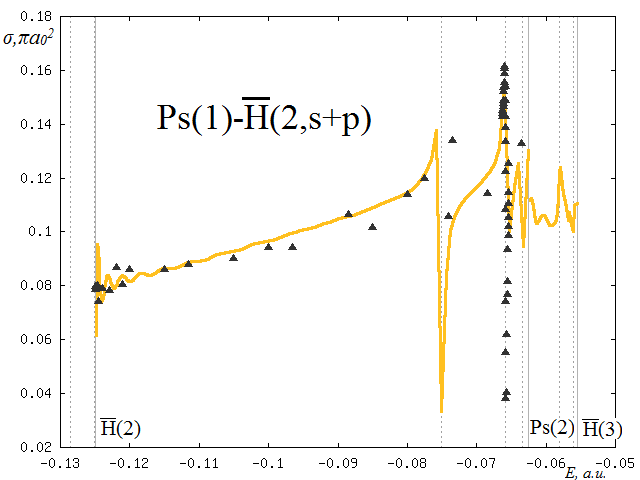}}
Ps(1)--$\bar{\mbox{H}}(2)$
\end{minipage}
\hfill
\begin{minipage}[h]{0.32\linewidth}
\center{\includegraphics[width=0.99\textwidth]{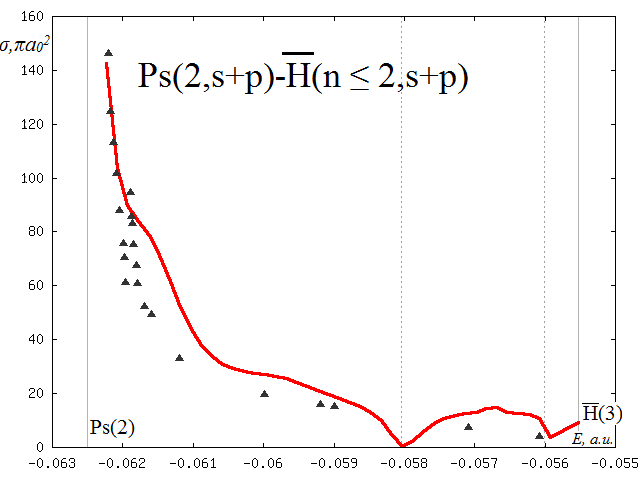}}\\
Ps(2)--$\bar{\mbox{H}}(n\le2)$
\end{minipage}
\caption{Antihydrogen formation cross sections. Black triangles mark points given in~\cite{lazau18}.}
\label{antiH}
\end{figure}

In the recent work~\cite{lazau18} cross sections for antihydrogen formation in the energy region between $\bar{\mbox{H}}(2)$ and $\bar{\mbox{H}}$(3) thresholds are studied in details.
For the comparison we plot our results on these cross sections in Fig.~\ref{antiH}.
The interest in~\cite{lazau18} was to find to what extent the resonances 
can enhance the antihydrogen formation cross sections.
Our energy resolution is better than that in~\cite{lazau18}, so our pictures may be useful for this purpose.
In both Figs.~\ref{CS_H} and~\ref{antiH} the resonances manifest themselves as peaks 
in some of the calculated cross sections.
Resonance energies found by different methods~\cite{ho04, varga08, yu12, umair14} are known with good accuracy.
In figures their positions are marked by vertical dashed lines.
All resonances are clearly seen in calculated cross sections, especially those of new processes which became possible at the previous threshold.
The cross sections of processes  $\mbox{Ps}(1,s)\rightarrow\mbox{Ps}(1,s)$ and $\bar{\mbox{H}}(2,s)\rightarrow\bar{\mbox{H}}(2,p)$ (and also $\bar{\mbox{H}}(2,s)\rightarrow\bar{\mbox{H}}(2,s)$ not shown in the figure) have 
sharp minima which look like resonances but do not coincide with any of known resonance positions. We agree with associating these minima in \cite{mitroy95} with the Ramsauer-Townsend effect.

\begin{figure}[t]
\begin{minipage}[h]{0.49\linewidth}
\center{\includegraphics[width=0.99\textwidth]{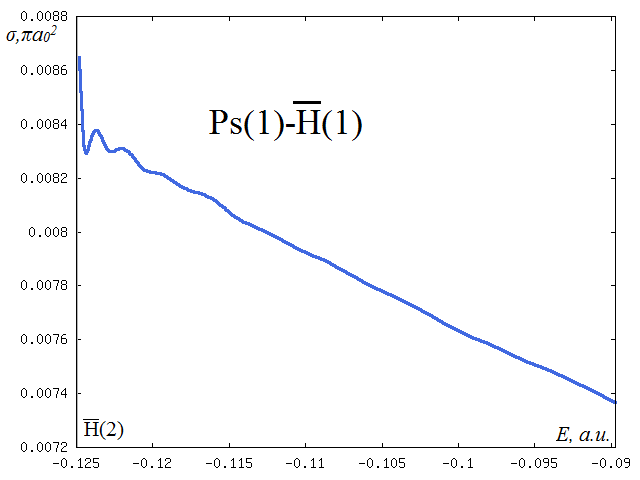}}\\
(a)
\end{minipage}
\hfill
\begin{minipage}[h]{0.49\linewidth}
\center{\includegraphics[width=0.99\textwidth]{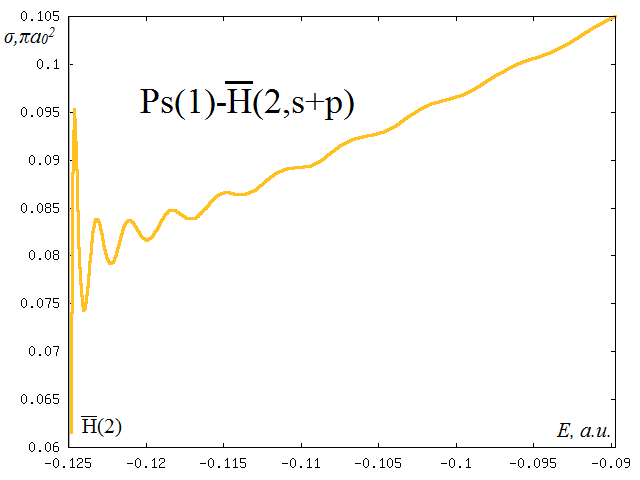}}
(b)
\end{minipage}
\begin{minipage}[h]{0.32\linewidth}
\center{\includegraphics[width=0.99\textwidth]{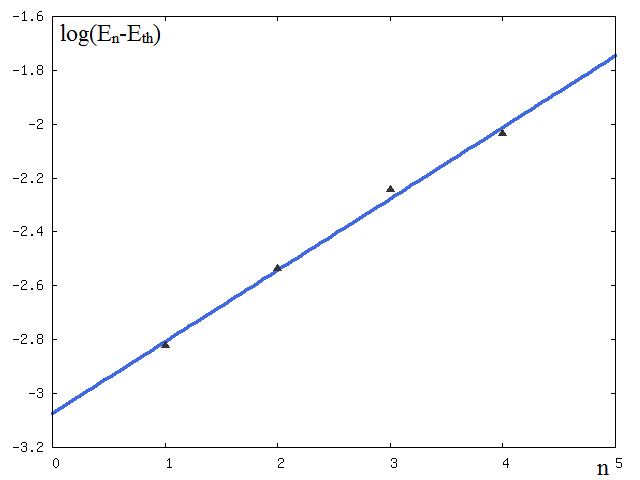}}\\
(c) Ps(1)--$\bar{\mbox{H}}$(1)
\end{minipage}
\hfill
\begin{minipage}[h]{0.32\linewidth}
\center{\includegraphics[width=0.99\textwidth]{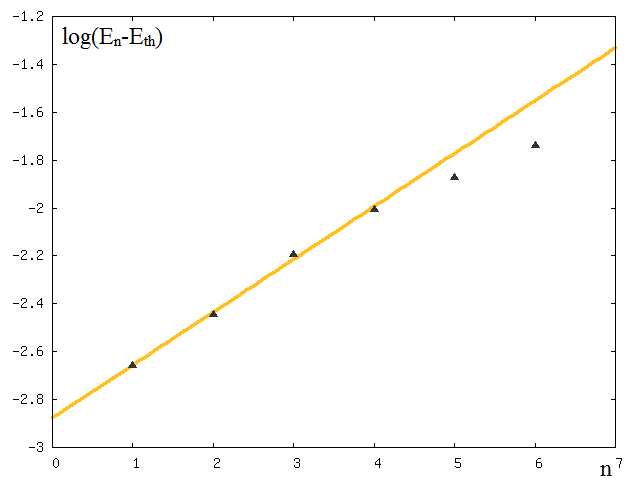}}\\
(d) Ps(1)--$\bar{\mbox{H}}$(2,s)
\end{minipage}
\hfill
\begin{minipage}[h]{0.32\linewidth}
\center{\includegraphics[width=0.99\textwidth]{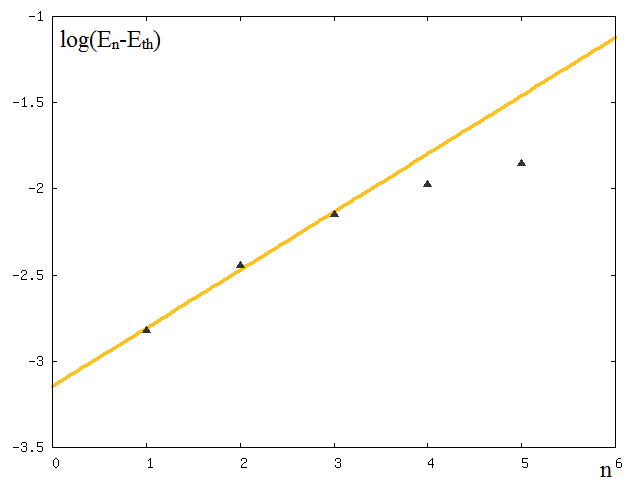}}\\
(e) Ps(1)--$\bar{\mbox{H}}$(2,p)
\end{minipage}
\caption{(a), (b) --- detailed plots of $\mbox{Ps}(1)-\bar{\mbox{H}}(1)$ and $\mbox{Ps}(1)-\bar{\mbox{H}}(2)$ cross sections in the energy region above the $\bar{\mbox{H}}(2)$ threshold. For these cross sections, the logarithm of the relative energy positions $\log(E_n-E_{th})$ of oscillations maxima with respect to their numbers $n$ are depicted in figures (c)--(e).}
\label{oscill}
\end{figure}

Special attention should be paid to just above the $\bar{\mbox{H}}(2)$ threshold oscillations of the cross sections in Fig.~\ref{antiH}.
We give more detailed plots of $\mbox{Ps}(1)\to\bar{\mbox{H}}(1)$ and $\mbox{Ps}(1)\to\bar{\mbox{H}}(2)$ cross section in the energy region above this $\bar{\mbox{H}}(2)$ threshold in Fig.~\ref{oscill}. Prominent oscillations of both cross sections and their character suggest to associate these oscillations with phenomenon predicted in \cite{gaili63, gaili63b}. According to \cite{gaili63, gaili63b} the energy position $E_n$ of the $n$th maximum of the oscillations must follow the rule
\begin{equation}
\log(E_n-E_{\mbox{th}})=A n + B,
\label{Grule}
\end{equation}
where $A$ and $B$ are constants and $E_{\mbox{th}}$ is the  threshold energy.  We plot the respective quantities for $\mbox{Ps}(1)\to \bar{\mbox{H}}(1)$ and $\mbox{Ps}(1)\to \bar{\mbox{H}}(2)$ near threshold oscillations in Fig.~\ref{oscill}.
As one can see the linear behaviour of $\log(E_n-E_{\mbox{th}})$ is near perfect in both cases of rearrangement cross sections except for last points. The latter can indicate the range of validity of approximations made in
\cite{gaili63, gaili63b} leading to (\ref{Grule}).
As for the behaviour of the $\mbox{Ps}(2)\to \bar{\mbox{H}}(n\le2)$ cross section on the right panel of Fig.~\ref{antiH}, we obviously cannot make such quantitative analysis of above the $\mbox{Ps}(2)$ threshold oscillations. Nevertheless we
can agree with \cite{lazau18} that there is an oscillation with the energy position close to $-0.06194\ a.u.$, which was also found earlier in \cite{Hu2002}.

\subsection{$e^+e^-\mbox{He}^{++}$ scattering}
Positron-positive helium ion is an example of positron--atomic target scattering 
in which asymptotic Coulomb interaction is present in one of the configurations.
There are a number of calculations in a wide energy region, among them are close coupling calculations using two center basis functions expansion~\cite{bransd01, kadyr18} and EFS-CDW method~\cite{jiao15}.
But to the best of our knowledge, there is lack of published results of calculation for the low-energy region.
In this work we have calculated K-matrices of all possible scattering processes in $e^+e^-\mbox{He}^{++}$ system in the total energy range from $-1.9997\ a.u.$ to $-0.12496\ a.u.$ with the energy step of calculation $0.0007\ a.u.$
In this interval elastic, excitations and rearrangement processes leading to He$^+$($n=1,2,3$) and Ps($n=1$)
atom states are possible.
The maximum linear size of K-matrix equals 7.
We proceeded as in the case of $e^+e^-\bar{p}$ system, using different sets of discretisation and cut-off function parameters for subsequent energy intervals between thresholds.
These intervals and corresponding parameters are given in Table~\ref{Hep_discr}.
We hold the accuracy of calculation  with errors not exceeding  1$\%$.

\begin{table}[t!]
\centering
\begin{tabular}{c|c|c|c}
\hline
 & He$^+$(1) -- He$^+$(2) & He$^+$(2) -- Ps(1) & Ps(1) -- He$^+$(4)\\
\hline
First component\\
\hline
$x_{\max}$ & 10.6 & 21.2 & 35.4 \\
$y_{\max}$ & 35.4  & 99 & 120 \\
$n_x$ & 60 & 60 & 90 \\
$n_y$ & 270 & 810 & 960 \\
$n_z$ & 15 & 27 & 21 \\
$x_0$ & 2.1 & 2.8 & 5.8 \\
\hline
Second component\\
\hline
$x_{\max}$ & 15 & 50 & 60 \\
$y_{\max}$ & 7.5  & 20 & 35 \\
$n_x$ & 60 & 90 & 105 \\
$n_y$ & 60 & 120 & 195 \\
$n_z$ & 15 & 18 & 42 \\
$x_0$ & 1 & 1 & 3.5 \\
\end{tabular}
\caption{Discretisation and cut-off function parameters used in calculations of $e^+e^-\mbox{He}^{++}$ system cross sections. The column headings denote the energy intervals associated with atoms excitations thresholds.}
\label{Hep_discr}
\end{table}

\begin{figure}[t]
\center{\includegraphics[width=0.99\textwidth]{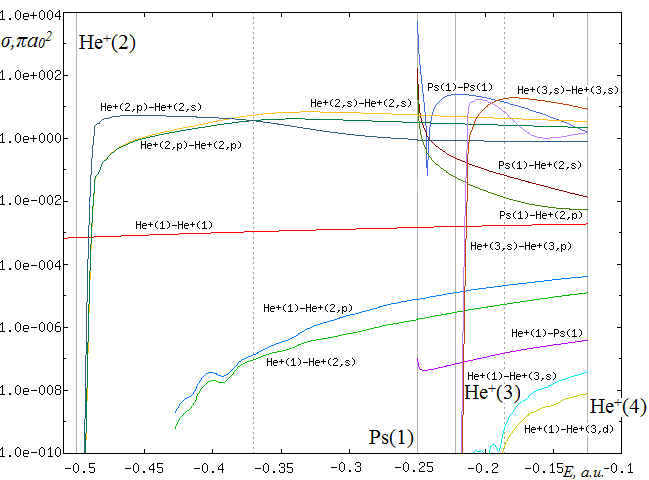}}
\caption{Cross sections in $e^-e^+\mbox{He}^+$ system. Vertical solid lines denote binary thresholds, vertical dashed lines mark resonance positions.}
\label{CS_Hep}

\end{figure}

The calculated values of cross sections are tabulated in Table~\ref{Hep_CS}.
Cross sections for the processes with positron or helium core colliding helium ion or positronium are presented in Fig.~\ref{CS_Hep}.
The largest cross sections are, as in the case of $e^+e^-\bar{H}$ system, the elastic ground state positronium cross section, and also that one which is associated with excited helium ion states.

\begin{table}[t!]
{\small
\begin{tabular}{c|c|c|c|c|c|c|c|c}
\backslashbox{}{$E$, a.u.} & 1.55 & 1.60 & 1.65 & 1.70 & 1.77 & 1.80 & 1.83 & 1.86 \\
\hline
$\sigma_{\mbox{He}^+(1)\rightarrow \mbox{He}^+(1)}$ & 0.000855 & 0.00101 & 0.00116 & 0.00133 & 0.00158 & 0.00168 & 0.00178 & 0.00188 \\
\hline
$\sigma_{\mbox{He}^+(1)\rightarrow \mbox{He}^+(2,s)}$ & $\sim$e-09 & $\sim$e-08 & 2e-07 & 6e-07 & 2.6e-06 & 4.4e-06 & 6.9e-06 & 1.1e-05\\
\hline
$\sigma_{\mbox{He}^+(1)\rightarrow \mbox{He}^+(2,p)}$ & $\sim$ e-10 & $\sim$e-08 & 3e-07 & 2.5e-06 & 1.1e-05 & 1.8e-05 & 2.6e-05 & 3.6e-05\\
\hline
$\sigma_{\mbox{He}^+(1)\rightarrow Ps(1)}$ & & & & & 1e-07 & 1e-07 & 2e-07 & 3e-07\\
\hline
$\sigma_{Ps(1)\rightarrow Ps(1)}$ & & & & & 20.6 & 19.6 & 8.82 & 3.00\\
\hline
$\sigma_{Ps(1)\rightarrow \mbox{He}^+(2,s)}$ & & & & & 0.366 & 0.102 & 0.0433 & 0.0199\\
\hline
$\sigma_{Ps(1)\rightarrow \mbox{He}^+(2,p)}$ & & & & & 0.0944 & 0.0214 & 0.00876 & 0.00584\\
\hline
$\sigma_{\mbox{He}^+(2,s)\rightarrow \mbox{He}^+(2,s)}$ & 1.12 & 3.35 & 6.64 & 6.63 & 5.11 & 4.59 & 4.10 & 3.66 \\
\hline
$\sigma_{\mbox{He}^+(2,p)\rightarrow \mbox{He}^+(2,s)}$ & 5.34 & 4.57 & 2.76 & 1.35 & 0.866 & 0.832 & 0.820 & 0.815\\
\hline
$\sigma_{\mbox{He}^+(3,s)\rightarrow \mbox{He}^+(3,s)}$ & & & & & & 9.87 & 18.4 & 11.7\\
\hline
$\sigma_{\mbox{He}^+(3,s)\rightarrow \mbox{He}^+(3,p)}$ & & & & & & 15.7 & 1.62 & 1.21\\
\hline
\end{tabular}
}
\caption{Scattering cross sections in $e^-e^+\bar{p}$ system (the energy is measured from the -1.9997 a.u. He$^+$(1) threshold)}
\label{Hep_CS}
\end{table}

All scattering cross sections associated with non excited atom states follow the well known laws of threshold behaviour~\cite{baz69}, which are presented briefly in Table~\ref{thresh} where $p$ means the relative momentum between
target and projectile.
The only exception is the cross section for $\mbox{He}^+(1)\to \mbox{Ps}(1)$ process.
At the $-0.25\ a.u.$  threshold, according to Table~\ref{thresh}, it should tend to zero linearly as $p\to 0$, but instead it grows up to some constant value.
This anomaly can be regarded as a sign of near threshold resonance, see the discussion below.
The near $\mbox{He}^+(2)$ threshold oscillations of cross sections for $\mbox{He}^+$  excitation to $\mbox{He}^+(2, s)$ and $\mbox{He}^+(2, p)$ reactions are evidently due to their very small values being disturbed by numerical errors.
The threshold behaviour of scattering cross sections associated with excited atom states is more complicated due to the degeneracy of energy levels and requires special treatment~
\cite{gaili63,gaili82,kvits89}.

\begin{table}[t!]
\begin{tabular}{c|c|c}
Process: & not charged & charged \\
\hline
elastic & const & $1/p^2$ \\
slow$\to$fast rearrangement & $1/p$ & $1/p^2$ \\
fast$\to$slow rearrangement & $p$ & const
\end{tabular}
\caption{Threshold behaviour of cross sections in the system of two compound particles $a$ and $X$, for elastic scattering and rearrangement to $b$ and $Y$. $p$ is the relative momentum of ingoing or outgoing slow particles, either oppositely charged or not charged, the angular momentum of the system equals zero.\cite{baz69}}
\label{thresh}
\end{table}

For $e^+e^-\mbox{He}^{++}$ system, resonance energies are worse known, there exist a number of discrepancies of results~\cite{igara97,igara04,han12} (and references therein).
Most authors agree that there are two broad resonances at $-0.371\ a.u.$ and $-0.188\ a.u.$~\cite{igara04} and one narrow resonance slightly below the positronium ground state formation threshold at $-0.250\ a.u.$~\cite{igara04,han12}.
These resonances positions are marked in Fig.~\ref{CS_Hep} by dashed vertical lines (dashed vertical line at
$-0.250\ a.u.$ almost coincides with vertical line denoting the positronium ground state threshold and is not visible).
We do not see the usual singular behaviour in the cross sections in the vicinity of the narrow resonance.
However, the discussed above anomalous threshold behaviour of the $\mbox{He}^+(1)\to \mbox{Ps}(1)$ process cross section at $\mbox{Ps}(1)$ threshold can indicate the presence of a resonance.
Broad resonances are not seen in the cross sections expectedly.

To check the existence of broad resonances we have used another approach based on the complex rotation method applied to the Schr\"odinger equation~\cite{yar98}.
We have found these broad resonances, their positions and widths are given in Table~\ref{res} and compared with results of~\cite{igara04}.

\begin{table}[t!]
\begin{tabular}{c|c|c}
Present work & (-0.3704, 0.1297) & (-0.1857, 0.0395) \\
\hline
\cite{igara04} & (-0.3705, 0.1294) & (-0.1856, 0.0393) \\
\end{tabular}
\caption{Broad resonance in the $e^+e^-\mbox{He}^{++}$ system energies ($E_r$, $\Gamma$) (in a.u.)}
\label{res}
\end{table}

The sharp local minimum is again seen in the $\mbox{Ps}(1)\to \mbox{Ps}(1)$ cross section for the direct process with neutral target. As above we assign this  minimum with the Ramsauer-Townsend effect.

\section{Conclusion}

In this paper the detailed calculations of low-energy reactive scattering in $e^-e^+{\bar p}$ and $e^+e^-\mbox{He}^{++}$ systems
for the zero total angular momentum have been performed with the use of the FM equations in total angular momentum representation. The total angular momentum representation for FM equations assumes all partial waves in subsystems to be included in the formalism, thus in this respect  the results obtained should be considered as complete.

The calculated cross sections in $e^-e^+\bar{p}$ system reproduce all known resonant peaks.
The sharp minima in elastic cross sections
${\bar p}-\mbox{Ps}(n=1)$ and $\mbox{He}^{++}-\mbox{Ps}(n=1)$ at low relative energies  display the  Ramsauer-Townsend effect. The Gailitis Damburg oscillations of the $\mbox{Ps}(n=1) \to \bar{\mbox{H}}(n=1)$ and $\mbox{Ps}(n=1) \to \bar{\mbox{H}}(n=2)$ cross sections just above the $\bar{\mbox H}(n=2)$ threshold are discovered
and the proper spacing with respect to the threshold of oscillation maxima are verified.

The two known  broad resonances \cite{igara04}  in $e^+e^-\mbox{He}^{++}$ system do not contribute into the cross section profile. The anomalous threshold behavior of the $\mbox{He}^+(n=1)\to \mbox{Ps}(n=1)$ cross section we suggest to explain by the existence of the narrow resonance found in~\cite{igara04,han12}.

\ack
This work was supported by Russian Foundation for Basic Research grant No. 18-02-00492. The calculations were performed on resources of the Computational Center of St Petersburg State University.

\section*{References}

\bibliography{Faddeev}

\end{document}